\documentclass{ws-procs975x65}
\usepackage[latin2]{inputenc}
\usepackage[IL2]{fontenc}
\usepackage[czech,british]{babel}

\newcommand{\dif}{\mathrm{d}}
\newcommand{\iu}{\relax\ifmmode{\mathrm{i}}\else\char"10\fi} 
\newcommand{\epow}[1]{\ensuremath{\eulern^{#1}}} 
\newcommand{\eulern}{\mathrm{e}} 

\begin{document}

\title{DYNAMICAL STABILITY OF FLUID SPHERES IN SPACETIMES WITH
  A~NONZERO COSMOLOGICAL CONSTANT%
  \footnote{This research has been supported by Czech grant MSM~4781305903.}}

\author{STANISLAV HLED\'IK$^{\dagger}$, ZDEN\v{E}K STUCHL\'IK$^\ddagger$
  and KRISTINA MR\'AZOV\'A$^\odot$}

\address{Institute of Physics, Faculty of Philosophy and Science,
  Silesian University in Opava, Bezru\v{c}ovo n\'{a}m. 13,
  Opava, CZ-746\,01, Czech Republic\\
  E-mail:
  $^{\dagger}$Stanislav.Hledik@fpf.slu.cz,
  $^{\ddagger}$Zdenek.Stuchlik@fpf.slu.cz,
  $^{\odot}$Kristina.Mrazova@fpf.slu.cz}

\begin{abstract}
  The Sturm--Liouville eigenvalue equation for eigenmodes of the radial
  oscillations is determined for spherically symmetric perfect fluid
  configurations in spacetimes with a nonzero cosmological constant and
  applied in the cases of configurations with uniform distribution of energy
  density and polytropic spheres. It is shown that a repulsive cosmological
  constant rises the critical adiabatic index and decreases the critical
  radius under which the dynamical instability occurs.
\end{abstract}


\bodymatter

\section{Introduction}\label{intro}

The internal Schwarzschild spacetimes with nonzero cosmological constant
($\Lambda\neq0$) and uniform distribution of energy density were given for
star-like configurations\cite{Stu:2000:ACTPS2:} and extended to more general
situations\cite{Boh:2004:GENRG2:}.  The polytropic and adiabatic spheres were
preliminary treated and
compared\cite{Stu-Hle:2001::GRPol,Hle-Stu-Mra:2003::GenRelPolCC}, neutron star
models with regions of nuclear matter described by different relativistic
equations of state that are matched were also
treated\cite{Urb-Mil-Sto:2005:RAGtime6and7:CrossRef}. Their stability can be
considered on energetic grounds\cite{Too:1964:ASTRJ2:} or it can be treated in
dynamical way\cite{Cha:1964:ASTRJ2:}. Here we determine the dynamical
stability conditions for the uniform density and polytropic spheres using the
approach of Misner et al.\cite{Mis-Tho-Whe:1973:Gra:}.

\section{Sturm--Liouville Equation}\label{SLeq}

Interior of a spherically symmetric configuration is described (in standard
Schwarzschild coordinates) by the line element
\begin{equation}
  \dif s^2 = -\epow{2\Phi}\,\dif t^2
             +\epow{2\Psi}\,\dif r^2
             +r^2(\dif\theta^2 + \sin^2\theta\,\dif\varphi^2)\,,
\end{equation}
with metric coefficients taken in the general form
\begin{equation}
  \Psi = \Psi(r,t)\,,\qquad\Phi = \Phi(r,t)\,.
\end{equation}
The perfect fluid distribution is given by energy density and pressure
profiles $\rho(r,t)$ and $p(r,t)$.  The static equilibrium configuration is
given by $\Phi_0(r)$, $\Psi_0(r)$, $\rho_0(r)$, $p_0(r)$ satisfying the Euler
equations. The radially pulsating configuration is then determined by
\begin{equation}
  q(r,t) = q_0(r) + \delta q(r,t),\qquad
  \delta q \equiv(\delta\Phi,\delta\Psi,\delta\rho,\delta p,\delta n)\,,
\end{equation}
where $n$ is the number density. The pulsation is given by the radial
displacement $\xi = \xi(r,t)$. The Euler perturbations $\delta q$ are related
to the Lagrangian perturbations measured by an observer who moves with the
pulsating fluid by the relation
\begin{equation}
  \Delta q (r,t) = q(r + \xi(r, t),t) - q_0 (r)
    \approx\delta q + q_0'\xi\,.
\end{equation}
Introducing a new variable $\zeta\equiv r^2 \epow{-\Phi_0}\xi$, the radial
pulsations are governed by
\begin{equation}
  W\ddot{\zeta}=\left(P\zeta'\right)' + Q\zeta
\end{equation}
with the functions $W(r)$, $P(r)$, $Q(r)$ determined for the equilibrium
configuration
\begin{gather}
  W \equiv (\rho_0+p_0)\frac{1}{r^2}\,\epow{3\Psi_0+\Phi_0}\,,\\
  P \equiv \gamma p_0 \frac{1}{r^2}\,\epow{\Psi_0+3\Phi_0}\,,\\
  Q \equiv \epow{\Psi_0+3\Phi_0}
    \left[\frac{(p_0')^2}{\rho_0+p_0}\frac{1}{r^2}
    -\frac{4p_0'}{r^3}-(\rho_0+p_0)
    \left(\frac{8\pi G}{c^4}p_0-\Lambda\right)
    \frac{\epow{2\Psi_0}}{r^2}\right]\,.
\end{gather}

The linear stability analysis can be realized by the standard assumption of
the displacement decomposition
\begin{equation}
  \zeta(r,t)=\zeta(r)\epow{\iu\omega t}\,.
\end{equation}
Then the dynamic equation reduces to the Sturm--Liouville equation and the
related boundary conditions in the form
\begin{gather}
  \left(P\zeta'\right)' + (Q+\omega^2W)\zeta = 0\,,             \label{eSL}\\
  \frac{\zeta}{r^3}\quad\text{finite as}\quad
  r \rightarrow 0\,,\qquad
  \gamma p_0 \frac{\epow{\Phi_0}}{r^2}\zeta' \rightarrow 0
  \quad\text{as}\quad r\rightarrow R\,.                         \label{e34}
\end{gather}
The Sturm--Liouville equation (\ref{eSL}) and the boundary conditions
determine eigenfrequencies $\omega_j$ and corresponding eigenmodes
$\zeta_i(r)$, where $i = 1,2,\ldots ,n$.
The eigenvalue Sturm--Liouville (SL) problem can be expressed in the
variational form\cite{Mis-Tho-Whe:1973:Gra:} as the extremal values of
\begin{equation}
  \omega^2 = \frac{\int_0^R\left(P\zeta'^2-Q\zeta^2\right)\,\dif r}%
                  {\int_0^R W\zeta^2\,\dif r}                   \label{e36}
\end{equation}
determine the eigenfrequencies $\omega_i$ and the corresponding functions
$\zeta_i (r)$ are the eigenfunctions that have to satisfy the orthogonality
relation
\begin{equation}
  \int_0^R \epow{3\Psi_0-\Phi_0}
    (p_0+\rho_0)r^2\xi^{(i)}\xi^{(j)}\,\dif r = 0\,.
\end{equation}

\section{Results and Conclusions}\label{rescon}

\begin{figure}[t]
\begin{center}
\psfig{file=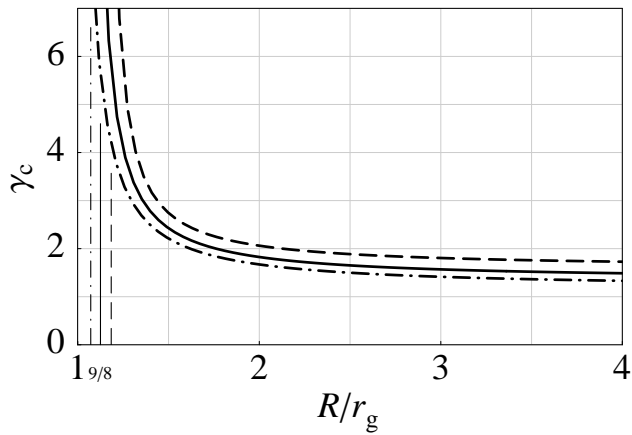,width=.65\linewidth}
\end{center}  
\caption{\label{Fig_1}Dependence of the critical value of adiabatic index
  $\gamma_{\mathrm{c}}$ on sphere radius $R$. \textit{Full curve:} vanishing
  cosmological constant $\lambda=0$; then $\gamma_{\mathrm{c}}$ diverges as
  $R\to9r_{\mathrm{g}}/8$ from above. \textit{Dashed curve:} positive
  cosmological constant $\lambda=0.1$, the point of divergence is shifted to
  $1.18421>9/8$, and $\gamma_{\mathrm{c},\lambda>0} >
  \gamma_{\mathrm{c},\lambda=0}$.  \textit{Dashed-dotted curve:} negative
  cosmological constant $\lambda=-0.1$, the point of divergence is shifted to
  $1.07143<9/8$, and $\gamma_{\mathrm{c},\lambda<0} <
  \gamma_{\mathrm{c},\lambda=0}$.}
\end{figure}

We applied the Sturm--Liouville approach to spheres with uniform energy
density\cite{Stu:2000:ACTPS2:}, and polytropic
spheres\cite{Stu-Hle:2001::GRPol}. The case of uniform spheres can be properly
taken as a test bed of the dynamical instability problem~--- although these
solutions of the Einstein equations are of rather artificial character, they
reflect quite well the basic properties of very compact
objects\cite{Gle:1988:PHYSR3:}.
The dependence of the critical value of adiabatic index $\gamma\equiv(\partial
\ln p/\partial \ln n)_{\!s} = (n/p)(\Delta p/\Delta n)$ on configuration
radius $R$ is for the uniform case given in Fig.\,\ref{Fig_1} (see also
\cite{Boh-Har:2005:PHYSR4:}). The polytropic case is treated in details by
Stuchl\'{\i}k and Hled\'{\i}k.\cite{Stu-Hle-2:2005:RAGtime6and7:CrossRef}

\end{document}